\documentclass[aps,prd,twocolumn]{revtex4-2}

\usepackage{orcidlink}

\usepackage{graphicx}
\usepackage[caption=false]{subfig}

\usepackage{amsmath}
\usepackage{amssymb}

\usepackage{xcolor}
\usepackage{hyperref}
\usepackage{soul}

\usepackage{tabularx}

\newcommand{\cF}{\mathcal{F}}
\newcommand{\solar}{_{\odot}}
\newcommand{\umin}{_{\text{min}}}
\newcommand{\umax}{_{\text{max}}}
\newcommand{\uminmax}{_{\text{min/max}}}
\newcommand{\uref}{_{\text{ref}}}
\newcommand{\ustart}{_{\text{start}}}
\newcommand{\ufinish}{_{\text{finish}}}
\newcommand{\urss}{_{\text{rss}}}
\newcommand{\ugw}{_{\text{GW}}}
\newcommand{\urot}{_{\text{rot}}}
\newcommand{\ugte}{_{\text{GTE}}}
\newcommand{\upw}{_{\text{PW}}}

\begin{document}

\title{Piecewise frequency model for searches for long-transient gravitational waves from young neutron stars}

\author{Benjamin Grace\,\orcidlink{0009-0009-9349-9317}}
\email{Benjamin.Grace@anu.edu.au}
\author{Karl Wette\,\orcidlink{0000-0002-4394-7179}}
\email{Karl.Wette@anu.edu.au}
\author{Susan M. Scott\,\orcidlink{0000-0002-9875-7700}}
\author{Ling Sun\,\orcidlink{0000-0001-7959-892X}}

\affiliation{OzGrav-ANU, Centre for Gravitational Astrophysics, Australian National University, Canberra ACT 2601, Australia}

\begin{abstract}
In this work we characterise the performance of a new search technique designed to be sensitive to the remnants of binary neutron star systems. Sensitivity estimates of the new method on simulated data are competitive against those of other work. Previous searches for a gravitational-wave signal from a possible neutron star remnant of the binary neutron star merger event GW170817 have focused on short ($<500$~s) and long duration (2.5~hr -- 8~day) signals. To date, no such post-merger signal has been detected. We introduce a new piecewise model which has the flexibility to accurately follow gravitational-wave signals which are rapidly evolving in frequency, such as those which may be emitted from young neutron stars born from binary neutron star mergers or supernovae. We investigate the sensitivity and computational cost of this piecewise model when used in a fully coherent 1800-second $\cF$-statistic search on simulated data containing possible signals from the GW170817 remnant. The sensitivity of the search using the piecewise model is determined using simulated data, with noise consistent with the LIGO second observing run. Across a 100--2000~Hz frequency band, the model achieves a peak sensitivity of $h\urss^{50\%} = 4.4 \times 10^{-23} \text{Hz}^{-1/2}$ at 200~Hz, competitive with other methods. The computational cost of conducting the search, over a bank of $1.1 \times 10^{12}$ templates, is estimated at 10 days running on 100 CPU's.
\end{abstract}

\maketitle

\section{Introduction}

Since the first gravitational-wave discovery, the rate of new gravitational-wave detections has steadily increased. At the end of the third observing run of the Advanced LIGO~\cite{LIGO2015-AdvLIG} and Virgo~\cite{Virg2015-AdVScnIntGrWDt}, over 90 gravitational-wave events had been recorded~\cite{Collaboration2021}. Of these events, two are binary neutron star (BNS) coalescences~\cite{Abbott2017, Abbott190425}, while the remaining detections are predominantly binary black hole coalescences with a number of possible binary neutron star-black hole events. With gravitational-wave detectors continuing to improve in sensitivity and detection range, and with more detectors coming online, it is expected that the catalog of events will increase and that new astrophysical sources will contribute to these discoveries. It is prudent to begin preparing for the challenges of detecting these new sources in anticipation of their discovery.

The gravitational-wave event GW170817 is of particular significance; not only is it the first binary neutron star detection, but its discovery was paired with electromagnetic counterparts~\cite{Abbott2017}. The counterparts allowed for precise sky localisation of the source, and the gravitational-wave signal provided profound insights into neutron star physics. Despite the significance of this event, the nature of the remnant of the binary system after coalescence is surrounded by much uncertainty. The general consensus is that the remnant object is now a black hole, but between coalescence and today, there was an unknown period of time when a neutron star might have been present. Some groups however have claimed a stable neutron star may have formed after the coalescence \cite{VanPutten2019}.

Four scenarios are typically considered for the evolution of the remnant object following a binary neutron star event. The remnant object after coalescence may: i) immediately collapse into a black hole; ii) form a hypermassive neutron star before collapsing into a black hole; iii) form a supramassive neutron star before collapsing into a black hole; or iv) form a stable long-lived neutron star~\cite{Ravi2014}. Hypermassive and supramassive neutron stars are expected to be born above the maximum allowed mass for a non-rotating neutron star. Hypermassive neutron stars are expected to support their weight through thermal pressure and differential rotation, with lifetimes on the order of milliseconds~\cite{Baumgarte2000,  Hotokezaka2013, Shapiro2000}, whereas supramassive neutron stars support their weight through rotation alone and may have lifetimes anywhere from seconds to hours~\cite{Ravi2014}. If a remnant neutron star is present, it is expected to have significant deformation and hence ellipticity, and be spinning at high frequencies~\cite{Ai2020}. Theoretical estimates of the maximum ellipticities range from $10^{-7}$--$10^{-4}$ for neutron stars born with large ($10^{15}$~G) magnetic fields~\cite{Colaiuda2008, Haskell2007a, Ciolfi2010}, while other approaches estimate maximum ellipticities between $10^{-8}$--$10^{-6}$~\cite{Ciolfi2010, Haskell2006, Ushomirsky2000, DeLillo2022}. One study has suggested that millisecond pulsars may have a minimum ellipticity of $10^{-9}$~\cite{Woan2018}.

Supernovae are another mechanism through which young neutron stars are born. Similarly to neutron stars born from BNS events, these neutron stars are expected to be spinning at high frequencies with significant deformations. Supernovae remnants provide ideal candidates for gravitational-wave follow-up as they have excellent sky localisation due to the electromagnetic event associated with them. This allows for directed gravitational-wave searches of these sources to be carried out, improving the chances of detection.

Gravitational-wave searches have been conducted to detect signals from a possible remnant of GW170817~\cite{TheLIGOScientificCollaboration2018, Abbott2019, Abbott2017b}. A claimed detection of a multi-second gravitational-wave signal following GW170817 was made~\cite{VanPutten2019}, but is considered implausible due to energy constraints, as discussed in~\cite{OlivEtAl2019-MtcSEBSNDtGrvExEmG}. No other searches has made a plausible detection claim to date.

Continuous-wave search techniques are specifically adapted to be sensitive to signals originating from long-lived neutron stars. Such systems are expected to be giving off nearly monochromatic signals. This, however, is not the case for young neutron stars, which are expected to be born spinning extremely rapidly and spinning down over shorter periods of time. The potential gravitational-wave signal from such stars is typically referred to as a \emph{long transient}. As the frequency of the gravitational waves emitted by the neutron star is proportional to the star's rotational frequency, and therefore also evolves rapidly, standard continuous-wave techniques are not suitable for detecting long-transient signals.

In this work, we present a new signal model which allows for continuous wave techniques to be adapted to long-transient signals from young neutron stars. In Section~\ref{Background}, we introduce a search framework commonly used for continuous waves and summarise prior searches for post-merger gravitational waves from GW170817. In Section~\ref{PiecewiseModel}, we describe the newly invented piecewise model. In Section~\ref{Implementation}, we describe how the piecewise model has been implemented for a long-transient search, and in Section~\ref{Performance}, we present the performance and sensitivity of our new method. Finally, in Section~\ref{Conclusion}, we summarise our results and outline the next steps for employing the piecewise model in a search for sources of long-transient waves.

\section{Background} \label{Background}

\subsection{Continuous-wave search framework}

For a rotating solid body, the power loss via gravitational-wave emission is given as~\cite{Ostriker1969}
\begin{align}
    P\ugw &= \frac{32 G}{5 c^{5}} I_{zz}^{2} \epsilon^{2} (\pi f)^{6}, \label{GWEnergy}
\end{align}
where $I_{zz}$ is the moment of inertia along the axis of rotation, $\epsilon$ is the ellipticity, $f$ is the gravitational-wave frequency (assumed here to be twice the rotational frequency of the neutron star), $G$ is the gravitational constant, and $c$ is the speed of light. The corresponding strain amplitude of the radiated gravitational waves is~\cite{Jaranowski1998}
\begin{align}
    h_{0} &= \frac{16\pi^{2}G}{c^{4}}\frac{\epsilon I_{zz}f^{2}}{D}, \label{GWStrain}
\end{align}
where $D$ is the distance from the detector to the source. The gravitational-wave signal $s$ at time $t$ at a detector is the linear superposition of four functions, $h_{i}$, which depend on the characteristic strain of the incoming gravitational wave given in Eq.~\eqref{GWStrain} and the antenna pattern of the detector~\cite{Prix2007, Jaranowski1998}:
\begin{equation}
    s(t, \mathcal{A}, \vec{\lambda}) = \sum_{i = 1}^{4} \mathcal{A}_{i} h_{i}(t, \vec{\lambda}).
\end{equation}
The parameters $\mathcal{A}_{i}$ are the canonical amplitudes of the wave, and are functions of the parameters $\phi_{0}$, $\psi$, $h_{0}$ and $\cos\iota$~\cite{Prix}. The first two of these parameters are the initial phase at $t = 0$ and polarisation angle of the gravitational wave, respectively. The parameter $\iota$ is the angle between the rotational axis of the source and the sky position vector $\vec{n}$. Note that $s$ is linear in the parameters $\mathcal{A}_{i}$. The vector $\vec{\lambda}$ is built from the phase parameters of the wave.  For an isolated neutron star of known sky position, $\vec{\lambda}$ is composed of the frequency, and derivatives of frequency in time, of the gravitational waves.

The $\cF$-statistic is the maximised log-likelihood ratio, $\ln\Lambda$. The likelihood ratio $\Lambda$ is the ratio of the probability of a signal being present in the data to the probability of no signal being present. It is defined by~\cite{Jaranowski1998}
\begin{align}
    \ln\Lambda(\mathcal{A}, \vec{\lambda}) &= \left(x | s\right) - \frac{1}{2}\left(s | s\right),
\end{align}
where $x$ is a continuous analog of noisy detector data, and $\left(\cdot | \cdot \right)$ is the scalar product
\begin{align}
    \left(x | y \right) &= \frac{2}{S}\int_{t\uref}^{t\uref+T}x(t)y(t)dt.
\end{align}
Here, $T$ is the length of data over which the search is conducted, $t\uref$ is a given reference time, and $S$ is the one-sided spectral density of the detector noise, assuming it is constant over time. The detection statistic $2\cF$ is commonly used in continuous gravitational-wave searches \cite{Jaranowski1998}. The maximisation over $\ln\Lambda$ to determine $2\cF$ is done analytically over the $\mathcal{A}$, given that the log-likelihood is linear in these parameters:
\begin{align}
2\cF(\vec{\lambda}) &= \max_{\mathcal{A}}\{ \ln\Lambda(\mathcal{A}, \vec{\lambda}) \}.
\end{align}
A search is then performed to find the optimal phase parameters $\vec{\lambda}$ which maximise $2\cF(\vec{\lambda})$. 

A search using the $\cF$-statistic will compute $2\cF$ for a set of phase parameters $\{\vec{\lambda}\}$. This set is known as the template bank. If a signal is present in the detector data, with parameters $\vec{\lambda_{S}}$, it is unlikely that it will coincide with any given $\vec{\lambda}$ within the template bank. It then follows that any signal recovered using a given $\vec{\lambda}$ will have some loss of signal-to-noise ratio, $\rho(\mathcal{A}, \vec{\lambda}_{S}, \vec{\lambda})$. The signal-to-noise ratio can then be used to define this mismatch between signal parameters $\vec\lambda_{S}$ and $\vec{\lambda}$~\cite{Prix2007}:
\begin{align}
\mu &= \frac{\rho^{2}(\mathcal{A}, \vec{\lambda}_{S}, \vec{\lambda}_{S}) - \rho^{2}(\mathcal{A}, \vec{\lambda}_{S}, \vec{\lambda})}{\rho^{2}(\mathcal{A}, \vec{\lambda}_{S}, \vec{\lambda}_{S})}, \label{Mismatch}
\end{align}
where $\rho^{2}(\mathcal{A}, \vec{\lambda}_{S}, \vec{\lambda}_{S})$ corresponds to the signal-to-noise ratio of a template perfectly matching to the signal. For templates $\vec{\lambda}$ close to the signal parameters $\vec{\lambda}_{S}$, such that $\Delta \vec{\lambda} = \vec{\lambda}_{S} - \vec{\lambda}$ is small, a second-order Taylor expansion of Eq.~\eqref{Mismatch} leads to the parameter space metric $\mathbf{g}$~\cite{Owen1996-STmGrvWInsBnCTmS, Astone2002, Prix2007}:
\begin{align}
\mu &\approx \Delta \vec{\lambda}^{T} \frac{-1}{2\rho^{2}(\mathcal{A}, \vec{\lambda}_{S}, \vec{\lambda}_{S})}\frac{\partial \rho^{2}(\mathcal{A}, \vec{\lambda}_{S}, \vec{\lambda})}{\partial\vec{\lambda}}\bigg|_{\vec{\lambda} = \vec{\lambda}_{S}} \Delta\vec{\lambda} \label{MismatchSNR} \\
&= \Delta \vec{\lambda}^{T} \mathbf{g} \Delta\vec{\lambda}, \label{MismatchEllipse}
\end{align}
where $\cdot^{T}$ represents a matrix transpose.

A useful approximation for the metric is the phase metric, $\mathbf{g}_{\phi}$. The phase metric only depends on the phase parameters $\vec{\lambda}$, and is defined by~\cite{Brady1998, Prix2007}
\begin{equation}
\begin{split}
[\mathbf{g}_{\phi}]_{ij} &= \langle \partial_{i} \phi(t, \vec{\lambda})\partial_{j} \phi(t, \vec{\lambda})\rangle \\ &\quad - \langle\partial_{i} \phi(t, \vec{\lambda})\rangle\langle\partial_{j} \phi(t, \vec{\lambda})\rangle. 
\end{split}
\label{phase_metric}
\end{equation}
Here, the $\partial_{i}$ are the partial derivatives with respect to the $i$th parameter of the templates $\vec{\lambda}$. The $\langle \cdot \rangle$ are time averages, defined as
\begin{align}
    \left\langle x \right\rangle &= \frac{1}{T}\int_{t\uref}^{t\uref + T}x(t)dt.
\end{align}
The function $\phi(t, \vec{\lambda})$ describes the phase evolution of a gravitational-wave signal given the parameters $\vec{\lambda}$ as a function of time $t$. For a typical continuous gravitational-wave search, the phase is given as~\cite{Jaranowski1998}
\begin{align}
\phi(t, \vec{\lambda}) &= 2\pi \sum_{s = 0}^{S\umax} f^{(s)}\frac{(t - t\uref)^{s + 1}}{(s + 1)!} + 2\pi \frac{\vec{r}\cdot\vec{n}}{c}f\umax, \label{phasemodel}
\end{align}
where $\vec{r}$ is the position vector of the gravitational-wave detector with respect to the solar system barycentre, $\vec{n}$ is the unit vector pointing from the Solar System Barycentre (SSB) to the source and $f\umax$ is the maximum frequency of the gravitational wave over the search band. The phase model $\phi$ is the time integral of a given gravitational-wave frequency model $f\ugw(t, \vec{\lambda})$:
\begin{align}
\phi(t, \vec{\lambda}) &= 2\pi \int_{t\ustart}^{t\ustart + t}f\ugw(t', \vec{\lambda}) dt'. \label{phase_integral}
\end{align}
In the case of Eq.~\eqref{phase_integral}, $f\ugw$ is typically chosen to be a second or third-order Taylor expansion \cite[e.g.][]{Abbott2022a, Owen2022, Abbott2022b}. If $f\ugw$ (and hence $\phi$) is linear in the parameters $\vec{\lambda}$, then the phase metric $\mathbf{g}_{\phi}$ will be constant.

If the metric is constant, Eq.~\eqref{MismatchEllipse} defines an ellipsoidal region around the point $\vec{\lambda}_{S}$. The maximum mismatch, $\mu\umax$, determines the size of this region. Geometrically, a template with a mismatch within $\mu\umax$ corresponds to a set of signal parameters falling inside one of the ellipsoids [as defined in Eq.~\eqref{MismatchEllipse}] centred on that template. Smaller values of $\mu\umax$ leads to smaller elliptical regions surrounding templates. If a set of signal parameters then falls within one of these ellipses, the value of $\mu$ computed from Eq.~\eqref{MismatchEllipse} will be reduced, implying a greater signal to noise ratio for the given template, by Eq.~\eqref{MismatchSNR}. Thus, a lower $\mu\umax$ may lead to greater search sensitivity, however it increases the computational cost of the search as reducing the size of the sensitive elliptical regions means more templates are required to cover the parameter space~\cite{Wette2014, Prix2007-TmpSrGrvWEfLCFPrS}. The value $\mu\umax$ is then typically chosen to optimise the search sensitivity within the limits of available computational resources. Values of $\mu\umax$ commonly range between 0.1 and 0.2. The computational cost of carrying out a continuous-wave search in this manner naturally scales with the template bank size.

Constructing the template bank using the fewest templates is a covering problem using ellipsoids over the parameter space, for which efficient algorithms using lattices exist~\cite{Wette2014, Prix2007-TmpSrGrvWEfLCFPrS}. Using these algorithms optimises the computational cost of conducting a search using the $\cF$-statistic. This optimisation is only possible if the phase model is linear in its own parameters.
The size of the template bank is approximated by~\cite{Jaranowski2012}
\begin{align}
    \mathcal{N} &= \theta \mu\umax^{-n/2} \mathcal{V} \sqrt{\det(\mathbf{g}_{\phi})}, \label{TBankEstimate}\\
    \theta(n) &= \sqrt{n+1}\left(\frac{n(n+2)}{12(n+1)}\right)^{n/2},
\end{align}
where $\theta$ is the normalised thickness of the template bank lattice, $n$ is the number of dimensions of the parameter space, $\mathcal{V}$ is the volume of the parameter space. Of the terms in Eq.~\eqref{TBankEstimate}, only the metric $\mathbf{g}_{\phi}$ and $\mathcal{V}$ depend on $T$, and one usually assumes that $\mathcal{V}$ scales only weakly with $T$. As the phase metric in Eq.~\eqref{phase_metric} is solely dependent upon the signal model and its parameters, then so too is the template bank size in Eq.~\eqref{TBankEstimate}.

For a continuous gravitational-wave search directed at a single sky position~\cite[e.g.][]{Abbott2022a, Owen2022}, the gravitational-wave frequency $f\ugw(t, \vec{\lambda})$ is defined by a Taylor expansion in two to three phase parameters: the gravitational-wave frequency and its first and second spin-downs (derivatives in time):
\begin{gather}
\vec{\lambda} = \{f, \dot{f}, \ddot{f}\} , \\
f\ugw(t, \vec{\lambda}) = f + \dot{f} ( t - t\uref ) + \frac{1}{2} \ddot{f} ( t - t\uref )^2 . \label{eq:Taylor}
\end{gather}
The parameter space is then defined by the range over which these parameters extend.
The bounds on each parameter arise from the \emph{general torque equation} (GTE):
\begin{align}
\frac{df\ugw}{dt} &= -k f\ugw(t)^{n}, \label{GTE}
\end{align}
with solution
\begin{equation}
f\ugte\left(f_{0}, n, k, t\right) = f_{0}\big[ 1 + (n - 1) k t f_{0}^{n - 1} \big]^{\frac{1}{1 - n}}. \label{GTEsoln}
\end{equation}
The constant $k$ contains information on the physical properties of the neutron star and $n$ is the braking index. The value of the braking index indicates the dominant mechanism through which the neutron star is losing energy. Values of interest are $n = 3$ for energy losses through dipolar electromagnetic radiation, $n = 5$ through mass-quadrupole gravitational radiation, and $n = 7$ through gravitational-wave emission from a current quadrupole (i.e.\ $r$-modes). For a spinning neutron star with known frequency and first and second frequency time derivatives, the braking index is given by $n = \ddot{f}f/\dot{f}^{2}$. By restricting the spin-down parameters to fall within certain values of the braking index, the parameter space for a gravitational-wave search is defined by the inequalities~\cite{WettEtAl2008-SrGrvWvCssLI},
\begin{align}
f\umin \leq &f \leq f\umax \label{TExpBound1}, \\ 
- \frac{f}{\tau(n\umin - 1)} \leq &\Dot{f} \leq -\frac{f}{\tau(n\umax - 1 )} \label{TExpBound2}, \\ 
\frac{n\umin\Dot{f}^{2}}{f} \leq &\Ddot{f} \leq \frac{n\umax\Ddot{f}^{2}}{f}. \label{TExpBound3}
\end{align}
Here $n\uminmax$ are the range of braking indices that confine the gravitational-wave search parameter space; and $\tau$ is a characteristic age of the neutron star, related to the constant $k$. The values of $f\uminmax$ constrain the range of frequencies searched over.

\begin{figure}[h]
    \centering
    \includegraphics[width=\columnwidth]{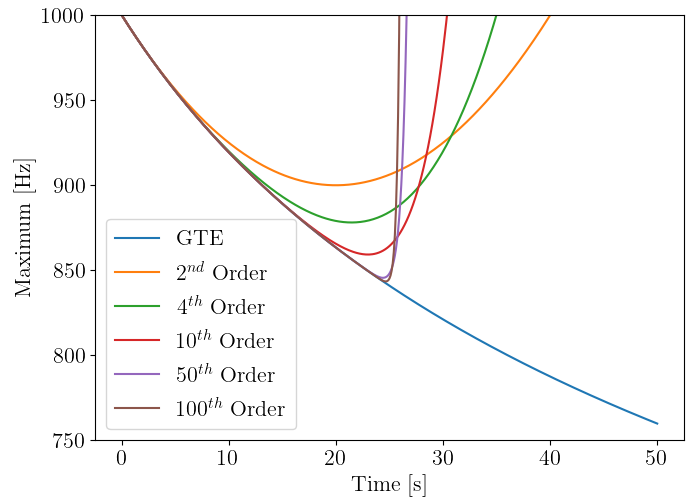}
    \caption{The expected spin down of a neutron star as modelled by Eq.\ \eqref{GTE} and its different order Taylor expansions taken from the point $t = 0$. Eq.\ \eqref{GTE} has parameters $n = 5$, $k = 10^{-14}$, $f_{0} = 1000$~Hz. The Taylor expansions diverge after 25s.}
    \label{fig:finite_convergence}
\end{figure}

A Taylor expansion signal model and the parameter space defined by the inequalities above are typically used for continuous-wave searches for isolated neutron stars. The frequencies of these sources are not expected to vary greatly over typical observation times of a year. For young neutron stars which are likely to be spinning down very rapidly over very short timescales, an example of which is shown in Fig. \ref{fig:finite_convergence}, many more spin-down parameters would be needed in order to track the evolution of the gravitational-wave frequency. For the example shown in Fig. \ref{fig:finite_convergence} 100 spin-down parameters are required to model the frequency of a young neutron star for 25~s. In extreme cases, for large values of $n$ and small values of $\tau$, Eq.~\eqref{GTE} has a finite interval of convergence for Taylor expansion approximations~\cite{Grace2018}. Figure \ref{fig:finite_convergence} shows an example of this finite interval of convergence of Eq. \eqref{GTE} for Taylor expansions of increasing orders. For long-transient searches for young neutron stars, therefore, new signal models must be considered.

\subsection{Prior post-merger searches for GW170817} \label{PrioSearches}

Previous searches carried out for a long-transient signal following the GW170817 merger have made use of both modelled and unmodelled methods~\cite{Abbott2017b, TheLIGOScientificCollaboration2018, Abbott2019}. Unmodelled searches are typically employed when searching for signals with unknown waveforms or large parameter space which would require a large template bank. Modelled searches alternatively use templates which possible signals may take the form of. These templates are then matched to data to calculate a detection statistic.

The STAMP~\cite{Thrane2011}, coherent WaveBurst (cWB)~\cite{Klimenko2008} and hidden Markov model tracking~\cite{Sun2019, Suvorova2016} methods are unmodelled search algorithms which have been applied to searching for a post-merger remnant of GW170817. STAMP uses spectrograms made from the cross-correlation of data between separated detectors and was designed for long-transient signals with durations of days to weeks~\cite{Thrane2011}. Pattern recognition algorithms are then applied to the STAMP spectrograms to determine detection statistics for potential candidate signals. The cWB algorithm operates by combining detector data coherently and, similar to STAMP, uses pattern recognition algorithms to identify candidate signals~\cite{Klimenko2008}. A search for a GW170817 remnant across a 1--4~kHz frequency band has been carried out in \cite{Abbott2017b} using both the STAMP and cWB methods. This search looked for short (1~s) and intermediate (500~s) duration signals reaching peak sensitivities of $2.1 \times 10^{-22}\text{~Hz}^{1/2}$ and $5.9 \times 10^{-22}\text{~Hz}^{1/2}$ respectively for 50\% confidence of detection.

Hidden Markov model tracking is a computationally efficient method based on a Markov chain, allowing for uncertainties in the signal frequency evolution model. The method operates by determining the probability for a hidden variable to transition from one state to another, such as how the gravitational-wave signal frequency may evolve from one frequency bin to another at each time step. The most probable sequence of transitions can then be determined using the Viterbi algorithm~\cite{Viterbi1967}. Searches using Hidden Markov model algorithms for a GW170817 remnant carried out in \cite{TheLIGOScientificCollaboration2018} were sensitive to a post-merger signal at a distance of 1~MPc. Compared to the distance of GW170817 of 40~Mpc this search was not sensitive to any plausible signals.

Modelled searches used for GW170817 remnant searches include variations on the Hough transform~\cite{Houg1959-McAnlBbChPct} such as the FrequencyHough~\cite{Antonucci2008} and Adaptive Transient Hough~\cite{Oliver2019} techniques. The FrequencyHough transform, for example, operates by mapping individual points from a frequency-time plane to lines on a frequency and spin-down plane. By mapping all points from the frequency-time plane to the frequency spin-down plane, lines accumulate and will intersect at points which correspond to signal parameters, if one exists~\cite{Krishnan2004}. The searches carried out in \cite{TheLIGOScientificCollaboration2018} also included the FrequencyHough and Adaptive Transient Hough algorithms. These algorithms were also found to be sensitive to a post-merger signal at a distance of 1~MPc, not capable of detecting a post-merger signal from GW170817.

\section{Piecewise model} \label{PiecewiseModel}

Young neutron stars born with large rotational frequencies are expected to spin down more rapidly than long-lived neutron stars. As a result, Taylor expansion models of the gravitational-wave frequency [Eq.~\eqref{eq:Taylor}] do not have sufficient accuracy to be used for a long-transient gravitational-wave search for young neutron stars. A piecewise model overcomes the shortcomings of a Taylor expansion: whenever a particular approximation to the gravitational-wave frequency begins to break down, a new piecewise segment can commence with a new approximation. By repeating this process, a piecewise model can in principle be used for a long-transient wave search over arbitrary observation times.

\begin{figure}
    \centering
    \includegraphics[width=\columnwidth]{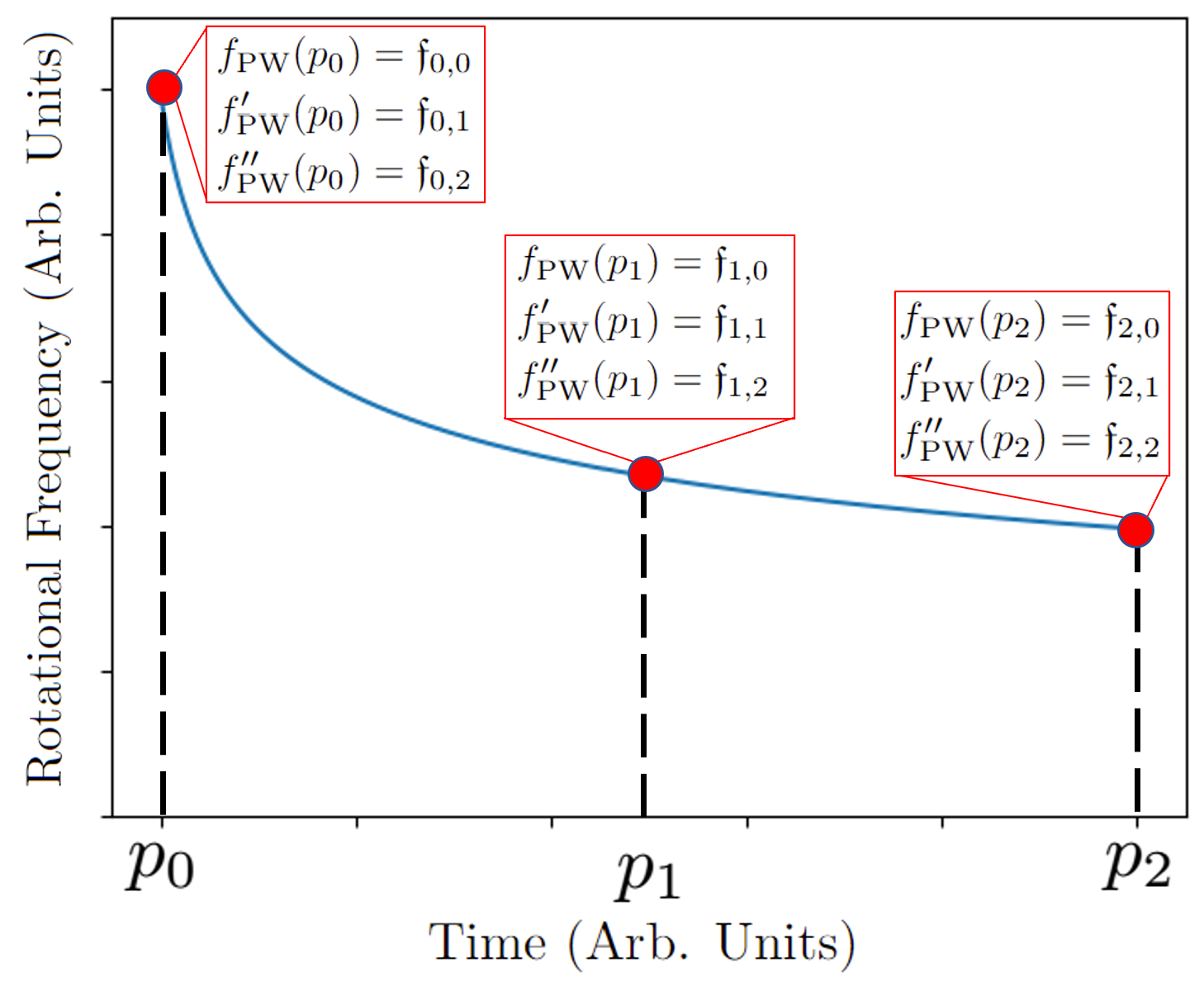}
    \caption{A visual representation of the piecewise model and its parameters. The piecewise model is equal to the value of its parameters at each knot at the appropriate derivative order.}
    \label{fig:PWVisual}
\end{figure}

The piecewise model proposed in this work models the gravitational-wave frequency as
\begin{align}
f\upw(t) &= \begin{cases} 
      f_{0}(t) & p_{0} \le t < p_{1}, \\
      f_{1}(t) & p_{1} \le t < p_{2}, \\
      ... & ... \\
      f_{N}(t) & p_{N} \le t \le p_{N + 1},
   \end{cases}  \label{eq:fupw} \\
\intertext{where}
f_{i}(t) &= \sum_{s=0}^{S - 1} \mathfrak{f}_{i,s} B^{0}_{i, s}\left(u_{i}(t)\right) + \mathfrak{f}_{i + 1, s}B^{1}_{i, s}\left(u_{i}(t)\right). \label{PWModel}
\end{align}
The model has the following components:
\begin{itemize}
    \item $p_{i}$ are the \emph{knots} of the piecewise function: the times where the model switches between piecewise segments.
    \item $S$ is the number of spin-down parameters included in the model. Since $S = 0$ denotes frequency, the highest derivative order parameter included in the model is $S - 1$.
    The dimensionality of the parameter space of the piecewise model scales as $S(N + 1)$, where $N$ is the number of piecewise segments.
    \item $\mathfrak{f}_{i, s}$ are the phase parameters of the model. Each parameter is a time derivative of frequency of order $s$. The subscript $i$ refers to the knot to which the parameter is attached.
    \item $B_{i, s}^{0/1}$ are the basis functions of the model. The subscripts $i, s$ refer to the piecewise segment, and the phase parameter derivative associated with the function, respectively. The superscript 0 denotes that the function is attached to the knot at the beginning of the segment; similarly, the superscript 1 denotes the end of the segment. Outside of this segment, the basis functions are undefined.
    \item $u_{i}(t): [p_{i}, p_{i+1}] \rightarrow [0, 1]$ is an arbitrary function which maps time over the $i$th segment to the unit interval. The $B_{i, s}^{0/1}$ use $u_{i}(t)$ as a basis function. In this work, $u_{i}(t) = (t - p_{i})/(p_{i + 1} - p_{i})$ is a linear map. For brevity, we write $B_{i, s}^{0/1}(u_{i}(t))$ as $B_{i, s}^{0/1}(t).$
\end{itemize}
Note that the piecewise model, given in Eq.~\eqref{eq:fupw}, is linear in its phase parameters $\mathfrak{f}_{i, s}$. Furthermore, on each piecewise segment, we enforce the following conditions:
\begin{equation}
\begin{split}
\frac{d^{s} f_{i}(t)}{dt^{s}}\bigg|_{t = p_{i}} &= \mathfrak{f}_{i, s} , \\ 
\frac{d^{s} f_{i}(t)}{dt^{s}}\bigg|_{t = p_{i + 1}} &= \mathfrak{f}_{i + 1, s}.
\end{split} \label{PWReq}
\end{equation}
With these conditions, the parameters $\mathfrak{f}_{i, s}$ are given a physical interpretation, represented visually in Fig.~\ref{fig:PWVisual}. The modelled gravitational-wave frequency $f\upw(t)$ (and its derivatives) are equal to the parameters at the knots with which they are associated. Conversely, each parameter $\mathfrak{f}_{i, s}$ represents the gravitational-wave frequency ($s=0$) or frequency time derivatives ($s>0$) in time at the $i$th knot.

\begin{figure*}
\includegraphics[width=\textwidth]{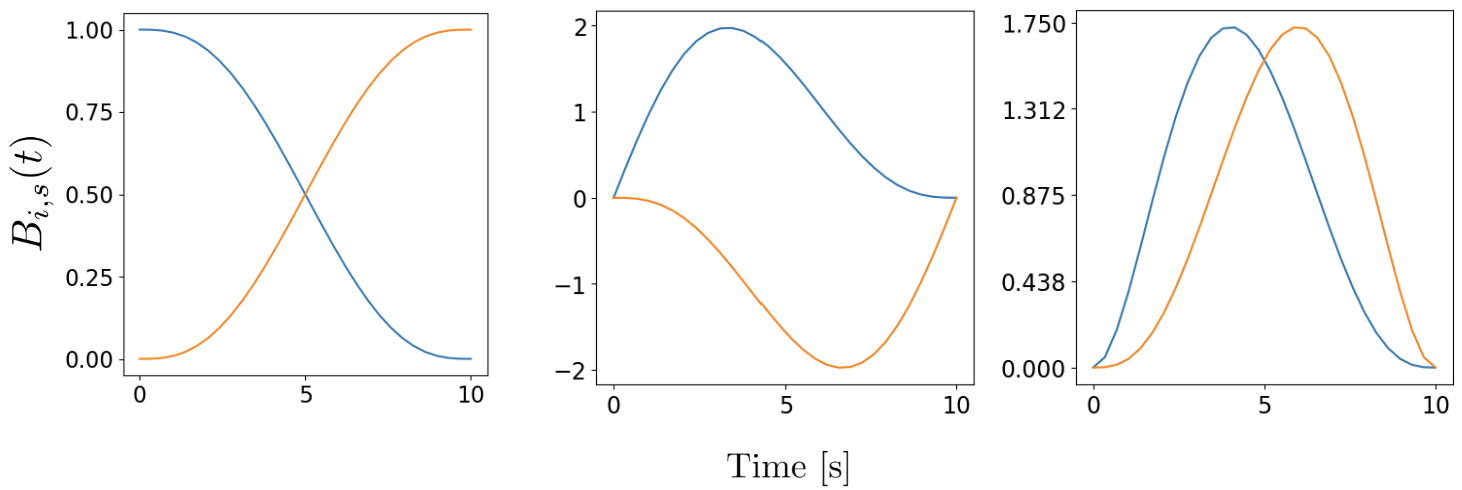}
    \caption{Six basis functions (left to right) $B_{i, 0}$, $B_{i, 1}$, and $B_{i, 2}$, for a piecewise segment where $p_{0} = 0$s, $p_{1} = 10$s and $S = 3$. The blue represents the $B_{i, s}^{0}$ basis functions while orange shows the $B_{i, s}^{1}$ basis functions. The piecewise model is a linear superposition of these functions.}
\label{BasisFunctionsEg}
\end{figure*}

Applying Eqs.~\eqref{PWReq} to Eq.~\eqref{PWModel} leads to the following conditions on the basis functions:
\begin{equation}
\begin{aligned}
\frac{d^{r}}{dt^{r}}B^{0}_{i, s}(p_{i}) &= \delta^{r}_{s}, &
\frac{d^{r}}{dt^{r}}B^{1}_{i, s}(p_{i}) &= 0, \\
\frac{d^{r}}{dt^{r}}B^{0}_{i, s}(p_{i + 1}) &= 0, &
\frac{d^{r}}{dt^{r}}B^{1}_{i, s}(p_{i + 1}) &= \delta^{r}_{s},
\end{aligned} \label{BasisFunctionReq}
\end{equation}
where $\delta^{r}_{s}$ is the Kronecker delta.
Beyond these conditions, we have complete freedom in how the basis functions are built. In this work, we set the basis functions to be polynomials of order $2S - 1$. Given that Eqs.~\eqref{BasisFunctionReq} are a linear system, they are easily solved for the polynomial coefficients. Figure~\ref{BasisFunctionsEg} shows the form of the basis functions for the case $S = 3$.

For the Taylor expansion signal model, the general torque equation [Eq.~\eqref{GTE}] is used to define the parameter space boundaries [Eqs.~\eqref{TExpBound1}--\eqref{TExpBound3}]. These bounds assume that the physical properties of the neutron star do not change over time; for example, the braking index for any template that satisfies Eqs.~\eqref{TExpBound1}--\eqref{TExpBound3} will be fixed throughout the search.
Young neutron stars, however, are expected to be evolving rapidly over short periods of time. As such, it is reasonable to assume that the braking index may evolve over the search, and therefore the bounds that are placed on the parameter space must accommodate this possibility. In addition, unlike the parameters of a Taylor expansion, the parameters of the piecewise model have a chronological order. The range of possible values for $\mathfrak{f}_{i, s}$ should therefore be influenced by its value at the previous knot, $\mathfrak{f}_{i - 1, s}$.

In this work, we use the solution to the GTE [Eq.~\eqref{GTEsoln}] to define the parameter space boundaries for the piecewise model. Unlike Eqs.~\eqref{TExpBound1}--\eqref{TExpBound3}, which use restrictions on the braking index and age of the source to inform these boundaries, we instead use Eq.~\eqref{GTEsoln} directly. The parameter space bounds for the piecewise model are:
\begin{align}
    \begin{split}
    \mathfrak{f}_{0, 0} &\geq  f\umin , \\
    \mathfrak{f}_{0, 0} &\leq f\umax ,
    \end{split}
    \label{f00_Condition} \\
    \begin{split}
    \mathfrak{f}_{i, 0} &\geq f\ugte(\mathfrak{f}_{i - 1, 0}, n\umax, k\umax, p_{i} - p_{i - 1}) , \\
    \mathfrak{f}_{i, 0} &\leq f\ugte(\mathfrak{f}_{i - 1, 0}, n\umin, k\umin, p_{i} - p_{i - 1}) ,
    \end{split}
    \label{fi0_Condition} \\
    \begin{split}
    \mathfrak{f}_{i, 1} &\geq f\ugte'(\mathfrak{f}_{i, 0}, n\umax, k\umax, 0) , \\
    \mathfrak{f}_{i, 1} &\leq f\ugte'(\mathfrak{f}_{i, 0}, n\umin, k\umin, 0) ,
    \end{split}
    \label{fi1_Condition} \\
    \begin{split}
    \mathfrak{f}_{i, 2} &\geq f\ugte''(\mathfrak{f}_{i, 0}, n\umin, k\umin, 0) , \\
    \mathfrak{f}_{i, 2} &\leq f\ugte''(\mathfrak{f}_{i, 0}, n\umax, k\umax, 0).
    \end{split}
    \label{fi2_Condition}
\end{align}
We have denoted the solution to Eq.~\eqref{GTE} as $f\ugte(f_{0}, n, k, t)$, where $f_{0}$ is the gravitational wave frequency at time $t = 0$, and $t$ is the time since the birth of the neutron star. The parameters $n\umin, n\umax, k\umin$ and $k\umax$ are predefined minimum and maximum values which the braking index $n$ and constant $k$ may range over.

Note that, in Eq. \eqref{fi0_Condition}, the bounds on $\mathfrak{f}_{i,0}$ are defined with respect to the piecewise frequency parameter on the previous knot, $\mathfrak{f}_{i - 1, 0}$, instead of on the first knot, $\mathfrak{f}_{0,0}$. As shown in Appendix \ref{GTEAppendix} the GTE satisfies
\begin{align}
f\ugte(F, n, k, T - t) = f\ugte(f_{0}, n, k, T),
\end{align}
where $F=f\ugte(f_{0}, t)$ and $T>t$. This property allows us to evolve the solution to the GTE forward from any given frequency and corresponding point in time, without needing to know its history. This allows Eq. \eqref{fi0_Condition} to be defined using only the frequency parameter on the previous knot.

The boundary condition Eq.~\eqref{f00_Condition} is the range of frequencies at time $t = 0$ we wish to search over. The boundaries defined by Eq.~\eqref{fi0_Condition} enforce all of the frequency parameters, and hence a given template, to follow the frequency evolution of Eq.~\eqref{GTE}. The conditions given in Eq.~\eqref{fi1_Condition} and Eq.~\eqref{fi2_Condition} for the first two spin-down parameters depend only on the frequency parameter which occurs at the same knot. These two conditions enforce that the braking index and $k$ value of the given template fall within the allowed ranges of $n$ and $k$.
Together, the boundary conditions Eqs.~\eqref{f00_Condition}--\eqref{fi2_Condition} define a parameter space where each template must follow the frequency evolution predicted by the GTE, and must always have a braking index and $k$ value which falls within a predefined range. This range of values is chosen by considering the physical properties of the source which is being targetted for a gravitational wave search.

The value of $k$ is highly uncertain, as it depends upon unknown neutron star physics such as the equation of state, magnetic field strength, and degree of physical deformation~\cite{Ostriker1969}. To estimate values of $k$ to define the parameter space, we equate the expression for the change in rotational energy of a solid body rotating at a frequency $f$ to a neutron star's dominant mode of energy loss. The resulting expression can then be rearranged into the form of the GTE to find an estimate of the value of $k$. The energy loss via gravitational-wave emission is given in Eq.~\eqref{GWEnergy}. The energy of a rotating solid body is
\begin{align}
    E\urot &= \frac{1}{2}\pi^{2} I_{zz} f^{2}.
\end{align}
If we assume that the star is losing energy only via gravitational-wave emission, we can equate $d E\urot/dt$ to $P\ugw$ in Eq.~\eqref{GWEnergy}:
\begin{equation}
\begin{split}
    \pi^{2}I_{zz}f\dot{f} &= \frac{32 G}{5 c^{5}} I_{zz}^{2} \epsilon^{2} (\pi f)^{6}, \\
    \dot{f} &= -\frac{32 G I_{zz} \pi^{4}\epsilon^{2}}{5 c^{5}} f^{5}.
\end{split}
 \label{GTE_Full_Form}
\end{equation}
The expression given in Eq.~\eqref{GTE_Full_Form} is of the same form as the GTE, which implies a value of
\begin{align}
    k &= \frac{32 G I_{zz} \pi^{4}\epsilon^{2}}{5 c^{5}}, \label{kGW}
\end{align}
with units of $\text{s}^{3}$. The value of $k$ for the case that the neutron star is emitting energy only via electromagnetic radiation may be found similarly, i.e.\ by equating $d E\urot/dt$ to the energy lost via an electric dipole~\cite{Ostriker1969}.

It is unlikely that a neutron star is undergoing energy loss via electromagnetic or gravitational wave radiation exclusively, so we do not expect the braking index for a neutron star to coincide exactly with the values of 3, 5 and 7. This is especially true considering the measured values of neutron stars which typically exhibit values of $n < 3$ \cite{DeAraujo2016}. It is therefore likely that the GTE does not encompass the complete physics of neutron star energy emission, however we use it to guide our assumptions on how a young neutron star will spin down. Values of the braking index outside of 3, 5 and 7 are interpreted to mean a mixing of energy loss mechanisms. For example, a value of $n = 4$ could be interpreted as energy loss via electromagnetic and gravitational radiation combined.

In this work, we assume the predominant mode of energy loss in the neutron star is via gravitational-wave emission. We assume a fiducial value for the moment of inertia and optimistic accepted values for the eccentricity \cite{Colaiuda2008, Haskell2007a, Ciolfi2010} of a neutron star, quoted in Table~\ref{GW170817Params}. To cover a range of possible $k$ values for our search, we set the minimum $k$ value to be 10\% of the maximum value $k\umax$, which in turn is given by Eq.~\eqref{kGW}. It is considered unlikely that a newborn neutron star would be spinning below 50~Hz (100~Hz gravitational wave frequency). We then only search for signals with a minimum frequency of 100~Hz. These values are presented in Table~\ref{GW170817Params}.

\begin{table}
    \begin{tabularx}{\columnwidth}{l@{\extracolsep{\fill}}l@{\extracolsep{\fill}}r}
         \hline \hline
         Parameter & Symbol & Value \\
         \hline
         Number of spin-downs & $S$ &  2 \\
         Minimum braking index & $n\umin$ & 2 \\
         Maximum braking index & $n\umax$ & 5 \\
         Minimum initial frequency & $f\umin$ & 100~Hz \\
         Maximum initial frequency & $f\umax$ & 2000~Hz \\
         Principal moment of inertia & $I_{zz}$ & $10^{38}$ $\text{kg m}^{2}$ \\
         Maximum ellipticity & $\epsilon$ & $10^{-4}$ \\
         Minimum $k$ value & $k\umin$ & $1.72 \times 10^{-20} \text{ s}^{3}$ \\
         Maximum $k$ value & $k\umax$ & $1.72 \times 10^{-19} \text{ s}^{3}$ \\
         Maximum mismatch & $\mu\umax$ & 0.2 \\
         Short Fourier Transform timebase & $T_{\text{SFT}}$ & 10~s  \\
         Knots & $p_{0}, p_{1}$ & 0, 1800~s  \\
         \hline \hline
    \end{tabularx}
    \caption{Default physical parameters of the GW170817 search. This search has been conducted coherently over the full half hour duration. Other search parameters pertaining to the GW170817 remnant such as sky position can be found in the discovery paper~\cite{Abbott2017}.}
    \label{GW170817Params}
\end{table}

\section{Implementation} \label{Implementation}

In this section, we outline how the piecewise model has been implemented for a long-transient search. The implementation is freely available as part of the gravitational-wave data analysis library \textsc{LALSuite}~\cite{lalsuite}. The piecewise search code is implemented in both the Python and C programming languages.
The search parameters associated with this search are listed in Table~\ref{GW170817Params}, however the code allows for these parameters to be changed with user input.

To use the piecewise model its knots must first be set. The knots may be chosen by the user, as is the case in this work, or alternatively an algorithm exists which determines the longest possible segments allowable while keeping the piecewise model accurate enough to be used in a search. This algorithm relies on knowing what the maximum allowable difference between a signal and its closest matching template can be which still allows for detection. For an $\cF$-statistic search, this difference must not exceed $\sim 1/T$. This error requirement arises from the Discrete Fourier Transform (DFT), for which individual Fourier components are separated by $\Delta f \sim 1/T$. If the maximum difference between the template and a signal sits below the error threshold $\sim 1/T$, the model and the signal will have a maximised overlap in data. A knot algorithm has been written which determines the greatest spacing between knots for which the error between a candidate signal and its corresponding closest template differs by no more than $\sim 1/(p_{i + 1} - p_{i})$.  The code written to calculate the knots using the knot algorithm is provided in the search code. As these piecewise segment lengths are the maximum allowed under the given error requirements, if the user is selecting the piecewise knots they should ensure that the resulting piecewise segment lengths do not exceed that calculated by the knot algorithm.

The parameter space bounds Eqs.~\eqref{f00_Condition}--\eqref{fi2_Condition} are then set, allowing for the template bank to be constructed by the algorithms in~\cite{Wette2014}. The parameter space metric has been calculated symbolically, allowing for rapid computation. This was achieved by determining the symbolic form of the basis functions from Eqs.~\eqref{BasisFunctionReq}. A symbolic expression for the gravitational-wave phase [Eq.~\eqref{phase_integral}] is then calculated by substituting in the symbolic basis functions as a part of the piecewise model of Eq.~\eqref{PWModel}. The symbolic expression for gravitational-wave phase is then used to calculate the parameter space metric in Eq.~\eqref{phase_metric}.

With the piecewise knots and metric set, a search can then be carried out. The search code implementation uses the \textsc{ComputeFstat} method from the \textsc{LALSuite} library to calculate the $\cF$-statistic for each template in the template bank. The code calculates $2\cF$ for each detector separately for vetoing purposes, as well as $2\cF$ for the combined detectors. The templates with the largest detection statistics are stored as well as those with the lowest mismatches. The implementation returns timing data and template counts for performance investigations. 

\begin{figure}
    \centering\includegraphics[width=\columnwidth]{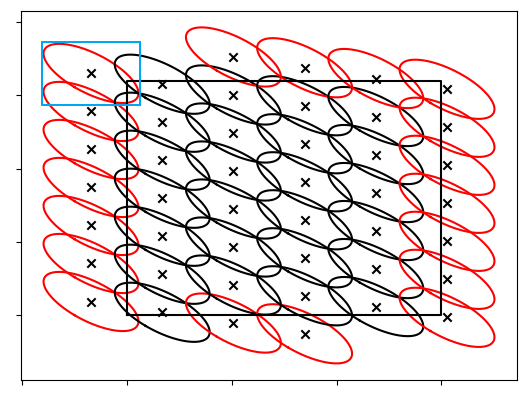}
    \caption{A simplified example of the lattice tiling used to cover a 2-dimensional rectangular parameter space. The parameter space is shown as a black rectangle, and templates are shown as crosses. Ellipses are those defined in Eq.~\eqref{MismatchEllipse} with their templates located at their centres. Red ellipses are associated with padding templates placed outside the parameter space. The padding templates are necessary for complete coverage of the parameter space. An example bounding box is shown in blue around the top left template.}
    \label{fig:PaddingExample}
\end{figure}

The implementation uses the template bank lattice algorithm of~\cite{Wette2014} for optimal computational cost. Additional templates placed outside the parameter space may be required in some instances to cover parts of the parameter space not covered by templates inside the space. Figure~\ref{fig:PaddingExample} illustrates when these additional \emph{padding} templates are required. For parameter spaces which are sufficiently narrow, a large portion of the space occurs close to the boundary. Without padding this leads to a significant percentage of the parameter space not being covered. The algorithm of~\cite{Wette2014} by default extends the parameter space by half of the metric \emph{bounding box} in order to add padding templates. The bounding box is the smallest hyperrectangle, with sides parallel to the parameter space coordinate axes, which encloses the parameter space metric ellipsoid [Eq.~\eqref{MismatchEllipse}]; an example is shown in Fig.~\ref{fig:PaddingExample}.

A search which uses multiple piecewise segments may require further padding considerations than discussed here. Multiple segments would naturally only be used for signals of a duration longer than the 1800s discussed in this work. Longer duration searches have a greater computational cost due to an increase in the time to compute the $\cF$-statistic and an increase in template bank size \cite{Wette2009}. The increase in template bank size is due to a finer template grid which results from using longer data segments \cite{Wette2014, Wette2009}. A finer grid would be more resilient to narrow parameter spaces however the computational cost may make a search at high frequencies unfeasible. This problem could be overcome by excluding the computationally expensive high-frequency bands from such a search, discussed in Section \ref{comp_cost}. A multiple-segment search has been considered for the supernova remnant 1987A, where a lower frequency band of 100--550~Hz has been considered. Lower frequency bands however are the regions in which additional padding is most needed. Further investigation into padding requirements is then necessary before a search on multiple segments is carried out. For searches which use multiple segments it is suggested that a semi-coherent search be used to reduce computational cost. Furthermore, for a semi-coherent search we propose that the piecewise segments are used as the individual search segments. The code for creating the mismatch histograms discussed here is provided in the search code.

\section{Performance} \label{Performance}

In this section, we characterise the behaviour and performance of the piecewise model implementation [Sec.~\ref{Implementation}]. We use the example of a follow-up search for gravitational waves from a post-merger remnant neutron star following GW170817. In~\ref{MetMismatchSection}, we investigate the template bank coverage of the parameter space, and identify regions excluded by the template covering. The computational cost of carrying out a search using the piecewise model is determined in~\ref{comp_cost} using two independent methods. The sensitivity of the method is estimated in~\ref{sensitivity} and compared to other searches.

To characterise the implementation, we apply it to synthetic detector data injected with simulated signals. These signals follow the same piecewise model as the search template; in this way, the mismatch of a template to the signal can be calculated using Eq.~\eqref{MismatchEllipse}. The parameter space of the piecewise model has been constructed to allow for the value of the braking index and k value to change over time. As such the set of signals possible within this parameter space is broad in scope. This parameter space is expected to encompass other traditionally considered continuous wave signals. As such, we do not expect injecting signals which use the piecewise model into data to lead to significant improvements in sensitivity estimates or systematic errors. Although we do not expect any systematic errors, testing the sensitivity of this method using different injected signal models would be a worthwhile test for future work.

To accurately simulate the decreasing amplitude of the signal expected from a young neutron star over short observations, the characteristic strain $h_0$ of the injected signals changes with time according to [cf. Eq.~\eqref{GWStrain}]
\begin{align}
h_0(t) &= h_{0}\left(\frac{f(t)}{f_{0}}\right)^{2}, \label{Signal_Strain}
\end{align}
where $h_{0}$ and $f_{0}$ are the initial characteristic strain and frequency of the signal, respectively, and $f(t)$ is the frequency of the signal at time $t$.

The synthetic detector data used in the simulations are in the form of 
Short Fourier Transforms (SFTs), a standard frequency-domain data product. They are generated using the \textsc{simulateCW} Python module of \textsc{LALSuite}~\cite{lalsuite}. Noise levels are chosen using noise curve data from the second observing run (O2) of the Livingston and Hanford detectors~\cite{Abbott2017b}. These noise levels have been chosen as the primary target for follow up using this method, GW170817, occurred in O2. It is expected that the sensitivity of the method will improve in subsequent runs as improvements in detectors are made. The parameters of the injected signals are chosen as random points within the parameter space of the piecewise model. The random points within the parameter space are chosen using the \textsc{RandomLatticeTilingPoints} method from \textsc{LALSuite}.

A single coherent segment of 1800~s  was selected, with only two knots at its start and end. This segment length is well below the maximum allowed by the knot algorithm (Sec.~\ref{Implementation}) which permits a maximum initial piecewise segment for the search parameters given in Table~\ref{GW170817Params} of 40,960~s. This configuration was selected as signal durations of approximately 1,800s after the GW170817 event have been mostly unexplored by other searches. A single coherent segment has been used to maximise search sensitivity and minimise the size of the parameter space which in turn minimises the computational cost. The SFT timebase is set to 10~s, giving 180 SFTs over the segment, to avoid the issues discussed in \cite{Covas2022}. 

All investigations discussed in this section are performed on 20 different frequency bands using the parameter values given in Table~\ref{GW170817Params}. The frequency bands all have their upper bounds $f\umax$ occurring at multiples of 100~Hz, beginning at 100~Hz and extending to 2000~Hz. The lower frequency bound for each band was determined by finding a frequency value which led to a template bank size of approximately $10^{6}$ for a reduced computational cost.

\subsection{Metric mismatch distributions} \label{MetMismatchSection}

\begin{figure}
\subfloat[]{\includegraphics[width=0.9\columnwidth]{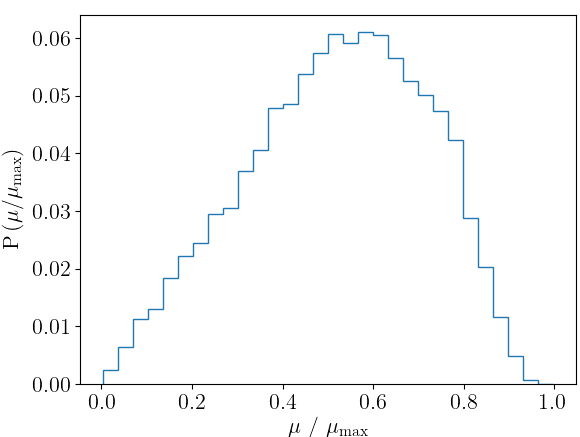}\label{fig:MH_fmax_1000}} \\
\subfloat[]{\includegraphics[width=0.9\columnwidth]{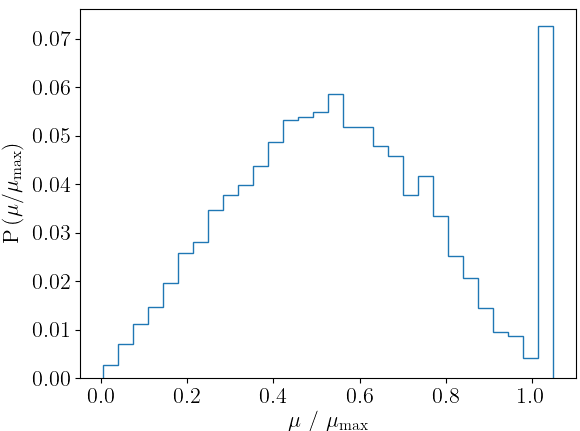}\label{fig:MH_fmax_100_No_Padding}} \\
\subfloat[]{\includegraphics[width=0.9\columnwidth]{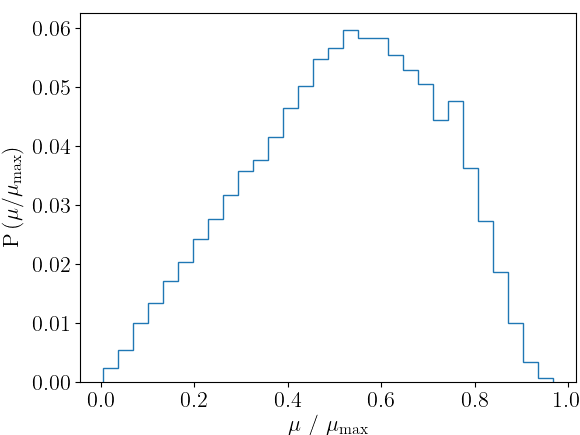}\label{fig:MH_fmax_100_Correct_Dist}} \\
\caption{\protect\subref{fig:MH_fmax_1000} Histogram of the minimum mismatch of $2\times10^{4}$ random searches in the frequency band 999.5-1000~Hz. \protect\subref{fig:MH_fmax_100_No_Padding} Histogram of the minimum mismatch of $2\times10^{4}$ random searches in the frequency band 92-100~Hz. The large number of mismatches above the maximum mismatch 0.2 occurs due to parameter space narrowing. \protect\subref{fig:MH_fmax_100_Correct_Dist} The same histogram as in \protect\subref{fig:MH_fmax_100_No_Padding} but now with appropriate padding added to the parameter space. The histogram now does not have any mismatches exceeding the maximal value.}
\label{fig:MH}
\end{figure}

\begin{figure}
\subfloat[]{\includegraphics[width=\columnwidth]{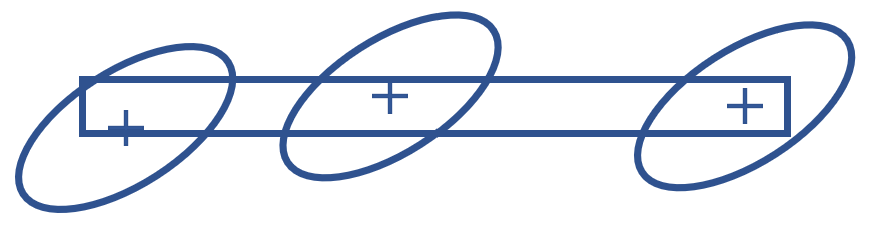}\label{fig:NarrowSpaceNoPadding}} \\
\subfloat[]{\includegraphics[width=\columnwidth]{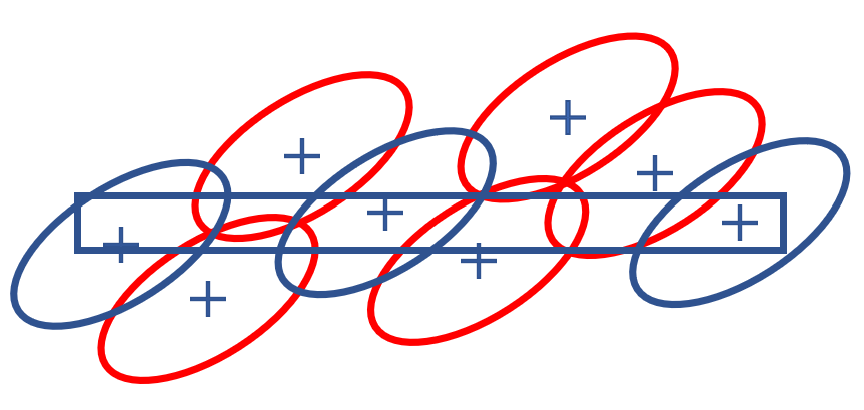}\label{fig:NarrowSpacePadding}}
\caption{\protect\subref{fig:NarrowSpaceNoPadding} A simplified narrow rectangular parameter space with its template covering. The tiling lattice is too coarse to cover a significant portion of the parameter space. \protect\subref{fig:NarrowSpacePadding} The same parameter space as \protect\subref{fig:NarrowSpaceNoPadding} with additional padded templates. Complete coverage is achieved however the number of padded templates is significant and outnumbers the templates found inside the parameter space.}
\label{fig:NarrowParameterSpaceExample}
\end{figure}

Mismatch distributions show whether the parameter space is covered completely by the template bank, and demonstrate the expected loss in signal-to-noise ratio of the search method. The mismatch distributions presented in this section are produced by carrying out $2\times10^{4}$ searches. In each search, the data contain a different injected signal, and the lowest mismatch found is stored. For complete coverage of the parameter space, all of the lowest mismatches should not exceed the maximal mismatch $\mu\umax$.

Figure~\ref{fig:MH_fmax_1000} shows the mismatch histogram for searches carried out at high frequencies. All mismatches sit below the maximal mismatch value and display the expected distribution; cf. Fig.~1 of~\cite{Wette2014}. At these higher frequencies, the template banks sufficiently cover the parameter space, and signals present within the space are recoverable.

Figure~\ref{fig:MH_fmax_100_No_Padding} shows the mismatch histogram for searches at lower frequencies. Here, a large percentage of searches have their lowest mismatch above the accepted maximal mismatch. At these lower frequencies, the parameter space has narrowed to the point that the default padding of half a bounding box is insufficient to cover the parameter space near the boundary. 
The parameter space is the most narrow in the dimension associated with the $\mathfrak{f}_{1, 1}$ parameter. The boundaries of the parameter space in this dimension are given by Eq.~\eqref{fi1_Condition}. In this dimension the parameter space of $\mathfrak{f}_{1, 1}$ narrows to the extent that it can fit between the now too-coarse template bank lattice. Figure~\ref{fig:NarrowParameterSpaceExample} shows a simplified example of how a sufficiently narrow parameter space fits between the template bank lattice, leaving a significant percentage of the space uncovered. This effect is remedied by including additional padding in the appropriate dimension, so that the parameter space is covered by the additional templates. For a search with parameter values given in Table~\ref{GW170817Params}, a single additional template above and below the minimum and maximum bounds of the $\mathfrak{f}_{1, 1}$ parameter is sufficient. 

Figure~\ref{fig:MH_fmax_100_Correct_Dist} shows the mismatch histogram for the same frequency band as Fig.~\ref{fig:MH_fmax_100_No_Padding} with the additional padding templates. With this correction, Fig.~\ref{fig:MH_fmax_100_Correct_Dist} follows the expected distribution, with all mismatches now below $\mu\umax$. For the parameter values shown in Table~\ref{GW170817Params}, additional padding is needed for all frequency bands with a maximum frequency below 550~Hz.

\subsection{Computational cost} \label{comp_cost}

The computational cost estimates in this section assume a follow-up search for a GW170817 post-merger remnant (with parameters given in Table~\ref{GW170817Params}) using the OzSTAR supercomputing cluster, which is planned for future work. We quote the computational cost of the search as the total time it would take for a 100 CPUs to complete the search across the full frequency band of 100 to 2000 Hz.

The computational cost for a coherent search is dominated by the speed at which $2\cF$ can be calculated for each template. The computational cost then scales linearly with the size of the template bank. In this subsection, we use two methods to estimate the computational cost of conducting a GW170817 post-merger search. The first method, the ``template estimate" relies on determining the total number of templates required by the search and multiplying this by the time taken to compute $2\cF$ for each template. The second method, the ``timing estimate" directly measures the time taken to complete a search on computationally inexpensive frequency bands across the parameter space. By interpolating the timing results from each smaller frequency band, an estimate on the total computational cost is achieved. Both methods rely on performing computationally inexpensive searches of narrow frequency bands across the 100--2000~Hz search range, and estimating the computational cost per unit frequency. Appendix \ref{CompCostAppendix} outlines the methods used for each computational cost estimate. The results of these estimates are presented in this subsection.

\begin{figure}
    \centering\includegraphics[width=\columnwidth]{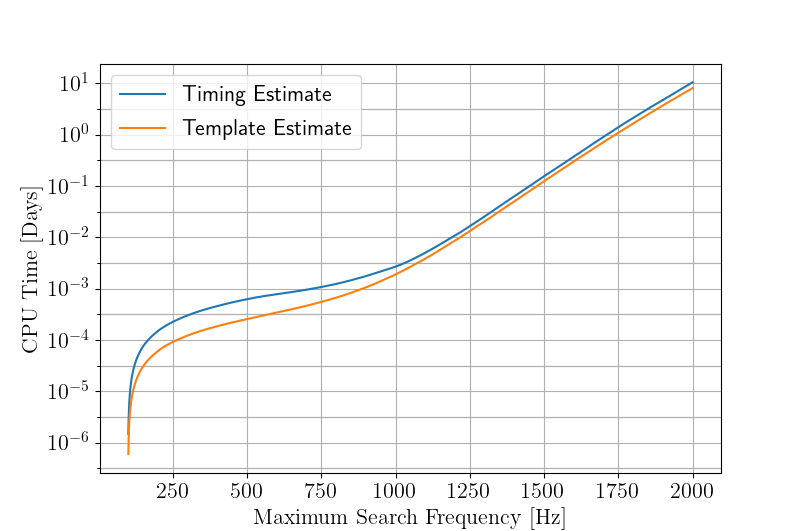}
    \caption{Estimated computational cost (for 100 CPUs) of conducting a piecewise search for the GW170817 remnant. The ``template estimate'' multiplies the expected size of the template bank by the time it takes to calculate $2\cF$ for each template.}
    \label{fig:comp_cost_fig}
\end{figure}

Figure \ref{fig:comp_cost_fig} presents the computational cost estimates for the search using two independent methods. Both estimates of the computational cost [Fig.~\ref{fig:comp_cost_fig}] are in broad agreement: $\sim 10$~days on 100 CPUs for a search covering 100--2000~Hz. The timing estimate is greater than the template count estimate, particularly below 1000~Hz; this is likely because the latter does not account for overheads of the search implementation that do not scale linearly with template bank size, i.e.\ tasks with a near-constant runtime. This is consistent with the relative discrepancy between the two estimates decreasing with frequency. While the timing estimate of the computational cost is therefore likely to be more reliable than the template estimate, the discrepancy is negligible for a maximum search frequency $\gtrsim 1000$~Hz. Errors in interpolating the convex curves in Fig.~\ref{fig:cost_per_freq} using linear interpolation may also lead to a slight overestimate of the computational cost.

\subsection{Sensitivity} \label{sensitivity}

To estimate the sensitivity of the search, we compute the \emph{detection probability} -- that a signal of a certain strain $h_0$ will be detected by this method -- as a function of $h_0$. To calculate the detection probabilities numerous searches on fake data with injected signals must be carried out. The number of searches, as well as the strength of injected signals varies for the different frequency bands for which the detection probabilities have been determined. Table \ref{DetProbTable} shows the set-ups used for calculating the detection probabilities in the different frequency bands. Different set-ups for determining the detection probabilities are used as the different frequency bands have different computational cost requirements. Higher frequency bands have a greater number of templates which require more time to carry out searches on. For this reason the higher frequency bands use less searches and investigate fewer values of $h_{0}$. 
Noise levels for all searches are chosen using O2 noise curve data from the Livingston and Hanford detectors~\cite{Abbott2017b}.

\begin{table}
    \begin{tabularx}{\columnwidth}{c c c c c c}
         \hline \hline
         $f\umin$ & $f\umax$ & $N_{Search}$ & $\log_{10}(\min(h_{0}))$ & $\log_{10}(\max(h_{0}))$ & $N_{h_{0}}$ \\
         \hline
         92      & 100  & 150 & -26 & -20 & 36 \\
         192     & 200  & 150 & -26 & -20 & 36 \\
         292     & 300  & 150 & -26 & -20 & 36 \\
         392     & 400  & 150 & -26 & -20 & 36 \\
         493     & 500  & 150 & -26 & -20 & 36 \\
         595     & 600  & 150 & -26 & -20 & 36 \\
         697     & 700  & 150 & -26 & -20 & 36 \\
         798     & 800  & 150 & -26 & -20 & 36 \\
         899     & 900  & 150 & -26 & -20 & 36 \\
         999.5   & 1000 & 150 & -26 & -20 & 36 \\
         1099    & 1100 & 100 & -26 & -20 & 36 \\
         1199    & 1200 & 100 & -24 & -20 & 24 \\
         1299.9  & 1300 & 100 & -24 & -20 & 24 \\
         1399.9  & 1400 & 100 & -24 & -20 & 24 \\
         1499.9  & 1500 & 100 & -24 & -20 & 24 \\
         1599.95 & 1600 & 100 & -24 & -21 & 18 \\
         1699.95 & 1700 & 100 & -24 & -21 & 18 \\
         1799.95 & 1800 & 100 & -23 & -20 & 14 \\
         1899.95 & 1900 & 100 & -23 & -21 & 12 \\
         1999.95 & 2000 & 100 & -23 & -21 & 12 \\
         \hline \hline
    \end{tabularx}
    \caption{The different configurations used for calculating the threshold statistic $2\cF^{*}$ and detection probabilities. $N_{Search}$ is the number of searches that were carried out on each frequency band. $N_{h_{0}}$ is the number of $h_{0}$ values investigated. These values were chosen using a logarithmic scale between the minimum and maximum $h_{0}$ values.}
    \label{DetProbTable}
\end{table}

\begin{figure}
    \centering\includegraphics[width=\columnwidth]{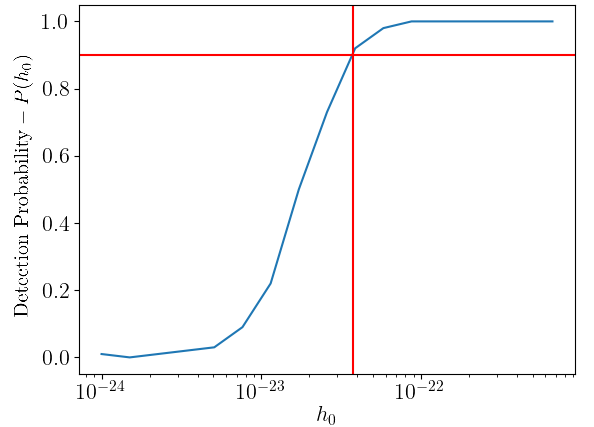}
    \caption{Detection probability as a function of $h_0$ for searches over the frequency band 1699.95--1700~Hz. The strain corresponding to a 90\% detection probability ($h_0 = 3.7 \times 10^{-23} \text{Hz}^{-1/2}$) is highlighted in red.}
    \label{fig:Det_Prob_Curve}
\end{figure}

We first calculate the threshold statistic $2\cF^{*}$. For the specific set up on each frequency band to calculate $2\cF^{*}$ and detection probabilities, see Table \ref{DetProbTable}. We perform a number of searches on data with no injected signal; for each search, the largest $2\cF$ is stored. The $2\cF$ occurring at the 99th percentile, corresponding to a 1\% false alarm rate, is selected as a threshold statistic. Another set of searches is then carried out, each with injected signals of a fixed strain $h_0$. Searches with a $2\cF$ above $2\cF^{*}$ are considered to have detected the injected signal. The fraction of searches where the injected signal is detected is the detection probability for the given $h_0$. Figure~\ref{fig:Det_Prob_Curve} plots an example detection probability curve at $\sim 1700$~Hz.

\begin{figure*}
    \centering
    \includegraphics[width=\textwidth]{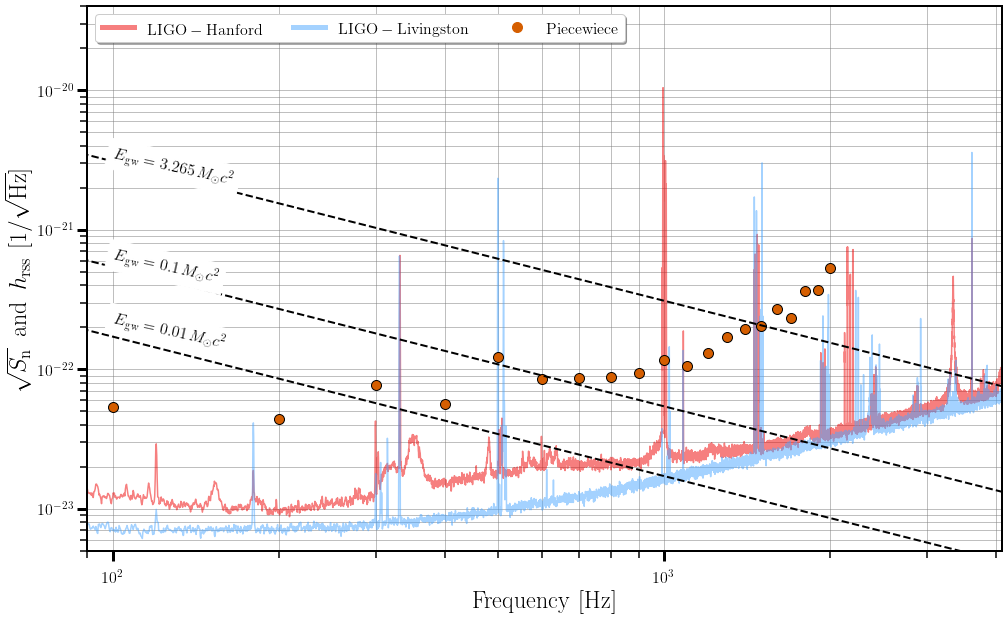}
    \caption{$h\urss^{50\%}$ sensitivity of the piecewise model, plotted as red circles. The noise amplitude spectral density ($\sqrt{S_{n}}$) of the LIGO Hanford and Livingston detectors during the O2 run are plotted in red and blue respectively. Lines of constant $E\ugw$ indicate the gravitational wave energy required for emission at a given $h\urss^{50\%}$ and frequency. Compare to Fig. 1 of~\cite{Abbott2017b}.} 
    \label{fig:sensitivity_curve}
\end{figure*}

Figure~\ref{fig:sensitivity_curve} presents the sensitivity estimates of the piecewise model. The sensitivities are expressed in terms of $h\urss^{50\%}$, the root sum squared strain amplitude corresponding to a 50\% detection probability~\cite{Abbott2019g}. Expressed in the time domain,
\begin{align}
    h\urss^{2} &= 2\int_{t\ustart}^{t\ufinish}\left(\left|\tilde{h}_{+}(t)\right|^{2} + \left|\tilde{h}_{\times}(t)\right|^{2}\right)dt,
\end{align}
where $t_{\text{start}}$ and $t\ufinish$ are the start and finish times, respectively, of the data used for the search. To calculate $h\urss^{50\%}$ for the piecewise model, we use strain values for $\tilde{h}_{+}$ and $\tilde{h}_{\times}$ corresponding to 50\% detection probabilities.

The piecewise model has a peak sensitivity of $h\urss^{50\%} = 4.4 \times 10^{-23} / \sqrt{\text{Hz}}$ at 200~Hz. It improves upon the sensitivities achieved in~\cite{Abbott2017b} by the STAMP searches by almost an order of magnitude, and is (at worst) within a factor of $\sim 2$ of the cWB method at high frequencies; see Section \ref{PrioSearches}. The $h\urss^{50\%}$ of the piecewise method increases with frequency in parallel with the detector noise, indicating a roughly constant signal-to-noise ratio of signals detectable by the method.

The energy emitted by isotropic gravitational waves from a source is given by~\cite{Sutton2013}
\begin{align}
    E\ugw^{\text{iso}} &= \frac{\pi^{2} c^{3}}{G}D^{2}\Bar{f}^{2} h\urss^{2}, \label{EmittedGWEnergy}
\end{align}
where $D$ is the distance to the source (in this case, GW170817). Figure~\ref{fig:sensitivity_curve} plots Eq.~\eqref{EmittedGWEnergy} at the most optimistic estimate of the energy available post-merger to be radiated in gravitational waves ($E\ugw = 3.265 M\solar c^2$) as well as at $0.1/3.265 \sim 3$\% and $0.01/3.265 \sim 0.3$\% of that estimate. The piecewise model would be sensitive to signals radiating $\gtrsim 3$\% of $3.265 M\solar c^2$ at frequencies $\lesssim 500$~Hz, and to signals radiating $\gtrsim 0.3$\% of $3.265 M\solar c^2$ at frequencies $\lesssim 200$~Hz. On the other hand, for frequencies $\gtrsim 1500$~Hz the sensitivity of the piecewise model is within an unphysical region where $E\ugw > 3.265 M\solar c^2$ would be required for a detectable signal.

The searches performed in~\cite{Abbott2017b} assumed several theoretical models of a post-merger neutron star: magnetars spinning down according to the GTE; secular bar-mode instabilities; and the post-merger component of simulated BNS merger waveforms. While the piecewise model demonstrates improved sensitivities to the first two (optimistic) models, it does not achieve the sensitivity required for the third (conservative) BNS merger simulation model. This model assumes BNS remnants emit $\lesssim 0.1 M\solar c^2$\% of energy in gravitational waves at $\sim 2000$~Hz; the piecewise model, in its current configuration, can only achieve such sensitivities at much lower frequencies $\lesssim 500$~Hz. Remnant neutron stars born in BNS mergers are not expected to be spinning at these frequencies \cite{Hotokezaka2013, Bauswein2012, Takami2014, Bernuzzi2015}.

\section{Conclusion} \label{Conclusion}

We have presented a new search technique for long-transient gravitational waves. It uses a piecewise model for the evolution of the gravitational-wave frequency with time, replacing the conventional Taylor series expansion used in searches for longer-lived continuous-wave signals. The parameters of the piecewise model have a clear physical interpretation, being the gravitational-wave frequency and the frequency derivatives at specific points in time. This physical interpretation of the parameters requires that the basis functions of the piecewise model satisfy certain criteria, while allowing for some freedom in choosing their time dependence; in this work, the basis functions are chosen to be polynomials in time. The piecewise model is then a linear superposition of these basis functions. We use the general torque equation to inform the boundaries of the search parameter space.

We examine the performance of the piecewise method assuming a search for a post-merger remnant of GW170817, using the $\cF$-statistic to search a frequency band of 100--2000~Hz for a fully coherent 1800-s signal. We consider the template bank size, estimated computational cost, and sensitivity of this search using the piecewise method. At frequencies below 550~Hz, the parameter space built from the GTE becomes narrow to the point that the template bank lattice is too coarse to completely cover the parameter space. Additional padding templates are added to address this issue. Further study for the template padding is required for searches which use multiple segments. A greater number of piecewise segments lends itself to finer template grids which are more resistant to narrow parameter spaces. Longer duration searches, however, come with greater computational cost and may require adjustment of the search frequency band. A semi-coherent search using the piecewise segments as the semi-coherent segments is suggested for multiple segment searches. Independent methods for estimating the computational cost of the search arrive at $\sim 10$ days on 100 CPUs of the OzSTAR supercomputing cluster to complete the search.

Sensitivity estimates in terms of the $h\urss^{50\%}$, the root sum squared strain at 50\% detection probability, are compared to past searches for a post-merger remnant of GW170817. The sensitivity of the piecewise method is competitive with past searches, and the 1800-s search duration complements past short ($< 500$~s) and long duration ($\gtrsim$~hr) searches for GW170817. With acceptable computational cost and competitive sensitivity estimates in hand, future work will look to perform a search for a post-merger remnant of GW170817 using the piecewise model.

While the piecewise method, as presented here, has been primarily motivated by the follow-up of binary neutron star merger events, the method is not fundamentally limited to these sources. Any long-transient or long-duration gravitational-wave sources which may have rapidly changing frequencies, beyond what conventional continuous-wave techniques are suited for, are appropriate for this method without alteration. One such source is the $\sim 36$-year old supernova remnant SN1987A, the youngest supernova remnant in the Milky Way. At that age, any SN1987A remnant neutron star is likely spinning at lower frequencies than expected for the neutron star remnants of binary neutron star mergers, and yet spinning down at a rate greater than that for which traditional continuous-wave techniques are suitable. The computational cost for the piecewise method is significantly reduced at lower frequencies, which would allow for longer data segments to be used for the piecewise model, increasing its sensitivity.

\section*{Acknowledgements}

This research was supported by the Australian Research Council under the ARC Centre of Excellence for Gravitational Wave Discovery, grant number CE170100004. B.G.\ would like to acknowledge the funding from the Australian Government Research Training Program (AGRTP) Scholarship for their research.
This work was performed on the OzSTAR national facility at Swinburne University of Technology. The OzSTAR program receives funding in part from the Astronomy National Collaborative Research Infrastructure Strategy (NCRIS) allocation provided by the Australian Government, and from the Victorian Higher Education State Investment Fund (VHESIF) provided by the Victorian Government.
This manuscript has document number LIGO-P2300314.


\appendix

\section{General Torque Equation} \label{GTEAppendix}
The solution to the GTE satisfies a convenient condition for determining the range of parameters.
For brevity, we write $f\ugte$ as a function of only $t$ and $f_{0}$. Let $f\ugte(f_{0}, t) = F$. Suppose that we want to determine the value of $f\ugte(F, T - t)$ with an initial frequency $F$ after a period of time $T - t$. Substituting into Eq.~\eqref{GTE}, we have
\begin{align}
    &f\ugte(F, T - t) \nonumber \\
    &= F\left(1 + (n - 1) k (T - t) F^{n - 1}\right)^{\frac{1}{1 - n}} \nonumber \\
    &= f\ugte(f_{0}, t)\left(1 + (n - 1) k (T - t) f\ugte(f_{0}, t)^{n - 1}\right)^{\frac{1}{1 - n}} \nonumber \\
    &= \left(f\ugte(f_{0}, t)^{1 - n} + (n - 1) k (T - t)\right)^{\frac{1}{1 - n}} \nonumber \\
    &= \left(f_{0}^{1 - n}\left(1 + (n - 1) k t f_{0}^{n - 1}\right) + (n - 1) k (T - t)\right)^{\frac{1}{1 - n}} \nonumber \\
    &= \left(f_{0}^{1 - n} + (n - 1) k t + (n - 1) k (T - t)\right)^{\frac{1}{1 - n}} \nonumber \\
    &= \left(f_{0}^{1 - n} + (n - 1) k T \right)^{\frac{1}{1 - n}} \nonumber \\
    &= f_{0}\left(1 + (n - 1) k T f_{0}^{n - 1}\right)^{\frac{1}{1 - n}} \nonumber \\
    &= f\ugte(f_{0}, T).
\end{align}

\section{Computational Cost} \label{CompCostAppendix}
\begin{figure}
    \centering
    \includegraphics[width=\columnwidth]{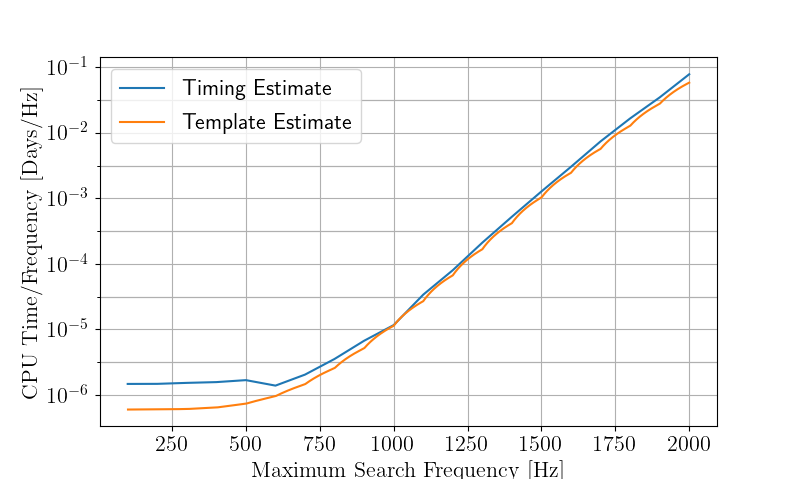}
    \caption{Estimated computational cost (for 100 CPUs) per unit frequency of the piecewise model. The ``template estimate'' multiplies the expected number of templates per unit Hz by the time it takes to calculate $2\cF$ for each template. The ``timing estimate'' interpolates the runtimes per unit Hz of a series of searches over small frequency bands.} 
    \label{fig:cost_per_freq}
\end{figure}

\begin{figure}
    \centering
    \includegraphics[width=\columnwidth]{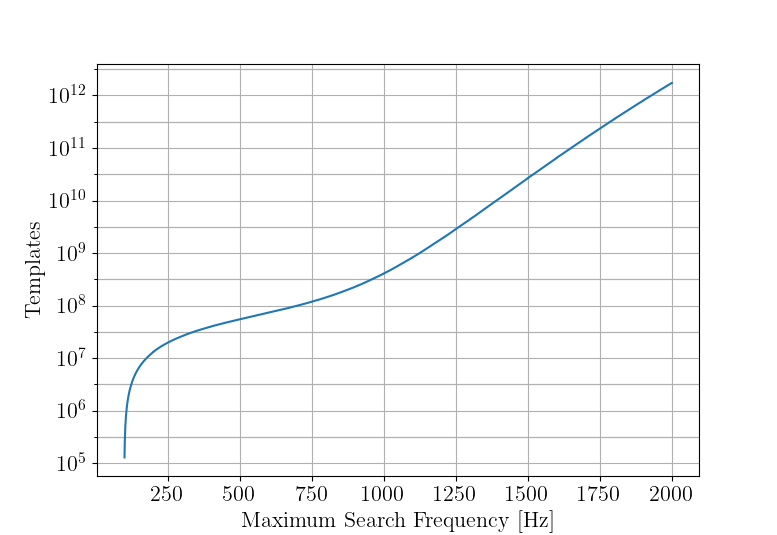}
    \caption{Estimated template bank size as a function of the maximum search frequency, with a minimum search frequency of 100~Hz.}
    \label{fig:temp_count}
\end{figure} 

The template estimate of the computational cost is achieved by first estimating the total number of templates contained within the template bank and then multiplying this number by the measured time taken to calculate $2\cF$ for an individual template. The time needed to calculate $2\cF$ for an individual template is measured directly, by timing a search on a template bank of known size. By repeating this measurement for a large number of different template banks at different frequency bands, we arrive at an averaged estimate of $2.5\times10^{4}$ templates per second on OzSTAR for a single CPU.

To estimate the size of the full template bank, we first directly count the size of 20 smaller template banks at 100Hz intervals from 100--2000~Hz. Each of these smaller template banks has the same parameters as in Table \ref{GW170817Params} except that each template bank has a smaller frequency band. Let each of these small template banks be labelled as $T_{i}$. The size of each frequency band has been chosen such that each $T_{i}$ contains approximately $10^{6}$ templates. We want to convert these template bank sizes into measurements of template bank size per unit frequency. If we have used frequency bands with widths of $\Delta f_{i}$, and each $T_{i}$ has a size $\mathcal{N}_{i}$ this gives us a measure of template bank size per unit frequency of $\mathcal{N}_{i}/\Delta f_{i}$.

To estimate the size of the full template bank, we need to integrate the values of template bank size per unit frequency across the full frequency band of 100--2000~Hz. From the 20 measurements of $\mathcal{N}_{i}/\Delta f_{i}$ we can linearly interpolate between these to achieve measurements of template bank size per unit frequency across the entire 100--2000~Hz range. This interpolation is done by
\begin{align}
    \frac{\mathcal{N}_{i}}{\Delta f_{i}} - \frac{f_{i}}{100\text{ Hz}}\left(\frac{\mathcal{N}_{i + 1}}{\Delta f_{i + 1}} - \frac{\mathcal{N}_{i}}{\Delta f_{i}}\right), \label{TBankInterpolation}
\end{align}
where the factor of 1/100~Hz arises from the interval between the smaller template banks. The result of this interpolation is shown in Fig.~\ref{fig:cost_per_freq}.

Finally, integrating \eqref{TBankInterpolation} gives us an estimate of the size of the template bank. Figure~\ref{fig:temp_count} presents the results of this integration as a function of the maximum search frequency. A GW170817 post-merger search [Table~\ref{GW170817Params}] across the full 100--2000~Hz frequency band requires $\sim 1.1 \times 10^{12}$ templates [Fig.~\ref{fig:temp_count}]. The higher frequencies have the greatest contribution to the template bank as a result of the larger parameter space volume in those regions, where the ranges given by Eqs.~\eqref{fi0_Condition} and~\eqref{fi1_Condition} are largest. As the parameter space metric is constant, the greater volume of the parameter space requires a greater number of templates. 
Figure~\ref{fig:comp_cost_fig} shows the computational cost as a function of maximum search frequency, found by multiplying the template bank size by the time taken to calculate $2\cF$ (i.e.\ $2.5\times10^{6}$ templates per second on OzSTAR for 100 CPUs).

The timing method of estimating the computation cost of the search directly measures the time taken to complete searches on small, inexpensive frequency bands (and using a single CPU). We then divide the runtime of each search by the width of the frequency band for each search; this yields estimates of runtime per unit Hz at discrete points. Linear interpolation between these estimates gives a curve of computational cost per unit Hz, shown in Figure~\ref{fig:cost_per_freq} as the ``timing estimate'' (and scaled for 100 CPUs). Integrating this curve across the frequency band then gives an estimate for the total computational cost, shown in Figure~\ref{fig:comp_cost_fig} as a function of the maximum search frequency.

\end{document}